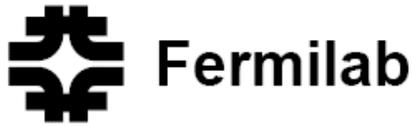

Fermilab-Conf-12-638-APC

December 2012

# HADRON PRODUCTION MODEL DEVELOPMENTS AND BENCHMARKING IN THE 0.7 – 12 GEV ENERGY REGION[*†]

N.V. Mokhov[#1], K.K. Gudima[2], S.I. Striganov[1]

[1]Fermilab, Batavia, IL 60510, U.S.A.
[2]Institute of Applied Physics, Academy of Sciences of Moldova

## Abstract

Driven by the needs of the intensity frontier projects with their Megawatt beams, e.g., ESS, FAIR and Project X, and their experiments, the event generators of the MARS15 code have been recently improved. After thorough analysis and benchmarking against data, including the newest ones by the HARP collaboration, both the exclusive and inclusive particle production models were further developed in the crucial for the above projects - but difficult from a theoretical standpoint - projectile energy region of 0.7 to 12 GeV. At these energies, modelling of prompt particle production in nucleon-nucleon and pion-nucleon inelastic reactions is now based on a combination of phase-space and isobar models. Other reactions are still modelled in the framework of the Quark-Gluon String Model. Pion, kaon and strange particle production and propagation in nuclear media are improved. For the alternative inclusive mode, experimental data on large-angle (> 20 degrees) pion production in hadron-nucleus interactions are parameterized in a broad energy range using a two-source model. It is mixed-and-matched with the native MARS model that successfully describes low-angle pion production data. Predictions of both new models are – in most cases - in a good agreement with experimental data obtained at CERN, JINR, LANL, BNL and KEK.

[*]Work supported by Fermi Research Alliance, LLC under contract No. DE-AC02-07CH11359 with the U.S. Department of Energy.
[†]Presented paper at the 11th Workshop on Shielding Aspects of Accelerators, Targets and Irradiation Facilities, SATIF-11, KEK, Tsukuba, Japan, September 11-13, 2012.
[#]mokhov@fnal.gov

**Introduction**

Fermilab, and US HEP in general, is moving to the Intensity Frontier, constructing/planning neutrino experiments (NOVA, LBNE, NuSTORM and Neutrino Factory), rare decay and high-precision experiments (Mu2e, g-2, ORKA, etc.) along with upgrades of the existing accelerators and planning for a Megawatt-scale multi-purpose Project X. Besides new challenges in the material and shielding aspects of these projects, all of the above requires reliable predictions of particle production at beam energies of 1 to 120 GeV, crucial for pions at $1 < E_p < 10$ GeV, the region where theoretical models traditionally have problems and existing experimental data contradict each other. The status of the models and recent developments of the MARS15 code event generators [1, 2] are described in this paper. The focus is on pion production in the difficult intermediate energy range – as a primary need of the above experiments – although new calculations for secondary protons and kaons are also shown.

**Issues with low-energy pion production**

General purpose particle transport codes such as Geant4, FLUKA and MARS15 do use event generators based on the intra-nuclear cascade models at energies below a few GeV and quark-parton models at higher energies. Both the groups have difficulties in the intermediate energy range of 1 to 10 GeV. To describe a low-energy particle production, a "formation length" should be introduced and determined from experimental data. To describe a large-angle particle production, interactions of a projectile with a multiple-particle bag should be taken into account. There are also difficulties specific to the models. In particular, descriptions of baryon resonance production cross-sections and interactions are needed, but these were never measured. At low projectile energies, an invariant mass of a chain (string) stretched between the quarks is so small that it is unclear how to transform it to real hadrons in a quark-parton model.

Experimental data on pion production are quite sparse. At small angles (< 10 degrees), only spectra of fast pions (p > 500 MeV/c) were measured. Backward pion production was studied in detail in Refs. [3-5]. There are several measurements of negative pion production in a broad angular range, but only for a limited number of target nuclei and few primary proton momenta [6-8]. Recently, the HARP collaboration has partially closed the gap between the small- and large-angle measurements. The large-angle study [9] covers pion angles from 20° to 123° and momenta from 0.1 to 0.7 GeV/c for primary proton and pion momenta 3, 5, 8 and 12 GeV/c. A part of the collaboration, HARP-CDP, has published similar data [10] based on a calibration different from that used by the main HARP collaboration. The cross sections measured are not that different for negative pions, but for the positive pions the difference between the HARP and HARP-CDP results reaches 60%.

The HARP collaboration performed comprehensive comparison of their large-angle data with Geant4 and MARS15 calculations [9]. The issues with both codes were revealed. The HARP-CDP group has also compared its measurements with the FLUKA and Geant4 simulations [11]. The Geant4 version 4.9.3 was found to not reproduce the energy dependence of pion production measured at the intermediate (20-50 degrees) and large (50-125 degrees) angles, while the FLUKA simulations agree with the data within 30%.

The HARP measurements [9] were fitted using a two-fireball approximation described in next section. The HARP cross sections were obtained by integration of this approximation into the HARP-CDP cuts. The HARP-CDP and HARP cross sections are presented in Fig. 1 in comparison with LAQGSM and FLUKA calculations. It is seen that the difference between results obtained by the two HARP groups is not large. The LAQGSM calculations agree with the data rather well. In a recent Geant4 version, the event generators were significantly improved and two new models, INCL and UrQMD, were included [12]. Results with UrQMD and FTFB are closer to the HARP-CDP data, while INCL produces results closer to the HARP data.

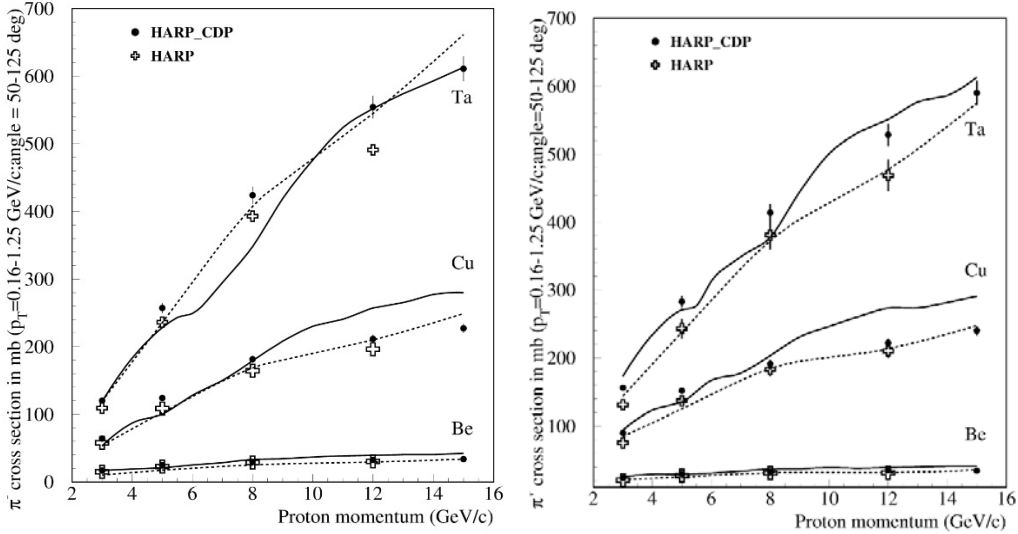

Fig. 1. $\pi^-$ (left) and $\pi^+$ (right) production cross-sections on Be, Cu and Ta nuclei *vs* proton momentum. Open symbols are approximation of HARP data [9], full symbols are HARP-CDP data [10], solid lines are FLUKA results from [11], and dotted lines are LAQGSM results.

**MARS15 inclusive pion production model development**

The main features of charged pion production at low and intermediate energies could be successfully described by a fireball model. In this model, the invariant cross-section of pion production reads

$$E\frac{d^3\sigma}{dp^3} \propto \exp(-\frac{T_{cm}}{T_0}) \propto \exp(-\frac{E - p\beta\cos(\theta)}{T_0\sqrt{1-\beta^2}}), \qquad (1)$$

where $T_{cm}$ is pion energy in the fireball rest-frame, $E, p, \theta$ are pion energy, momentum and angle in laboratory system, $\beta$ is fireball velocity. The low-statistic JINR bubble chamber data [6, 7] was fitted with a good $\chi^2$ by a relativistic form of the above formula ($E \approx p$). It turns out that to describe more precise measurements of negative pion production obtained by the KEK counter experiment for proton momenta of 3 and 4 GeV/c, one needs a two-source model [8]. At the same time, such a model fails to reproduce the LANL measurements [13] of pion production cross-sections at the proton energy of 730 MeV, especially at medium angles (15 – 60 degrees). To improve the model at such a low energy, one of the fireball sources can be changed to a term similar to the Fermi distribution. The resulting invariant cross-section can then be written as

$$E\frac{d^3\sigma}{dp^3} = p_1(1 + p_7\cos\theta)\exp(-\frac{T(1 - p_2\cos\theta)}{p_3}) + \frac{p_9(1 + p_8\cos\theta)}{1 + p_4\exp(T(1 - p_6\cos\theta)/p_5)} \qquad (2)$$

For $p_4 \gg 1$ this is similar to the relativistic two-fireball model [8], but now it fits the LANL data [13] with $\chi^2/n = 2.4$. Fig. 2 (left) illustrates the quality of the formula (2). It agrees with data [13] even better at larger angles.

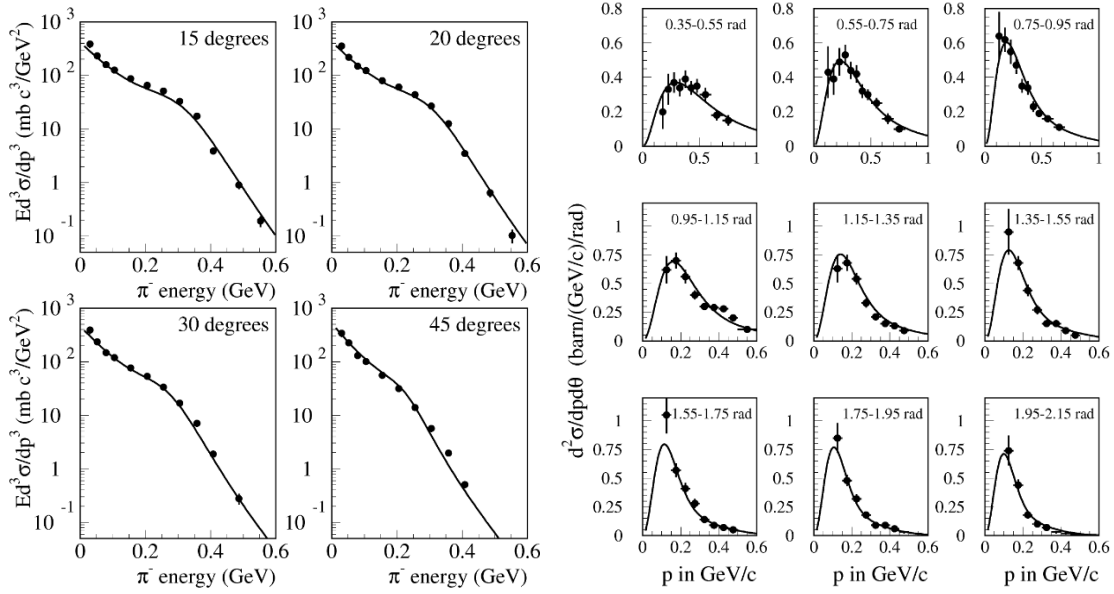

Fig. 2. $\pi^-$ production cross-sections in proton-lead interactions at 730 MeV (left) and 3 GeV/c (right). Solid curves are according to formula (2), symbols are experimental data [13] (left) and [9] (right).

Experimental data of the HARP collaboration [9] covers angles from 20° to 123°. This data is also successfully fitted by the two-source model (2) with $\chi^2/n \sim 1$ for all the HARP projectile momenta (3, 5, 8 and 12 GeV/c) and nuclei. A typical comparison is shown in Fig. 2 (right). The other measurements [3-5] at large angles (90, 119, 168 and 180 degrees) can also be included into the fitting procedure. The quality of the fit becomes slightly worse ($\chi^2/n \sim 2$) but is still quite acceptable. Fig. 3 (left) shows the comparison of the two-source model (2) with data at 180 degrees in a broad range of proton momenta. It is seen that the pion yield at 180 degrees grows with the proton momentum up to 5 GeV/c and remains constant after that for pions with a kinetic energy $\geq$ 150 MeV. The energy dependence of the negative pion spectra (normalized to the primary proton kinetic energy $T_0$) at the fixed angles is presented in Fig. 3 (right). The low-energy parts of the normalized spectra are independent of $T_0$, i.e. the yield of low-energy pions (< 150 MeV) grows with $T_0$ nearly linearly for $1 < T_0 < 7$ GeV.

The native inclusive MARS model [1] successfully describes the low-angle HARP data for primary protons with $3 < p_0 < 12$ GeV/c [14, 15]. A typical comparison for secondary pions and protons is presented in Fig. 4. A mix-and-match of the newly developed two-source description (2) and the native MARS model provides the complete description of pion production in proton- and pion-nucleus interactions.

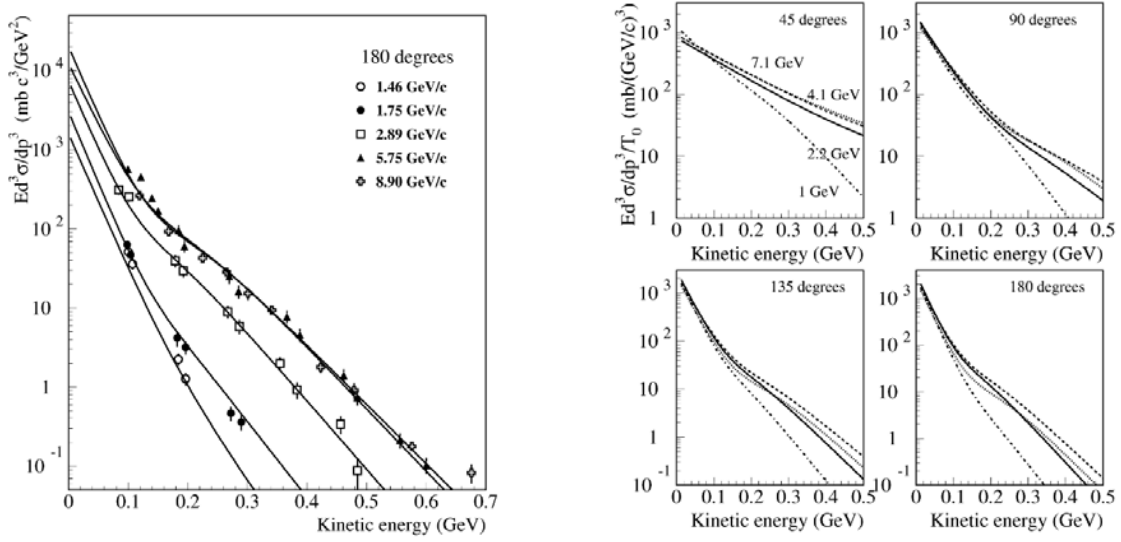

Fig. 3. $\pi^-$ production in proton-lead interactions as calculated according to formula (2) *vs* pion kinetic energy. Left: invariant cross-sections at the fixed angle of 180 degrees for proton momenta from 1.46 GeV/c to 8.9 GeV/c in comparison with data (see text). Right: invariant cross-sections (normalized to proton energy $T_0$) at various angles for $1 < T_0 < 7.1$ GeV.

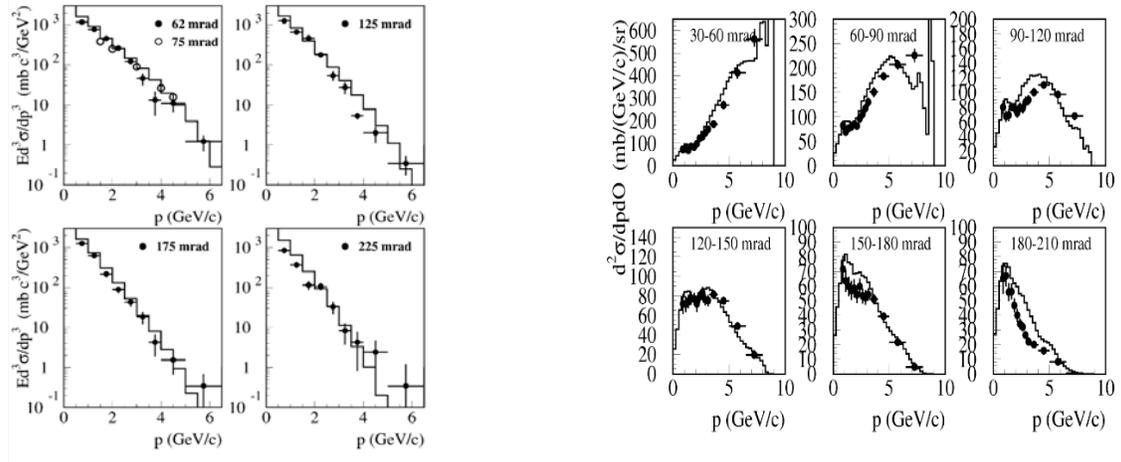

Fig. 4. MARS inclusive model *vs* 8 GeV/c HARP data [9]. Left: $\pi^-$ production cross-section at various angles in p+Ta interactions. Right: proton production differential cross-section at various angles in p+Be interactions.

**Quark-gluon string model developments**

The Quark-Gluon String Model, the LAQGSM code [2], is used in MARS15 for photon, hadron and heavy-ion projectiles at projectile energies from a few MeV/A to 1 TeV/A. This provides a power of full theoretically consistent modelling of exclusive and inclusive distributions of secondary particles, spallation, fission, and fragmentation products. The newest developments include: new and better approximations for elementary total, elastic, and inelastic cross sections for NN and πN interactions; several channels implemented for the explicit description with use of experimental data: N+N→N+N+mπ, π+N→N+mπ (m<5), B+B→B+Y+K, π+B→Y+K, π+B→B+K+Kbar, B+B→B+B+K+Kbar, Kbar+B→Y+π, and K+Kbar, N+Nbar pair production for the cms energy $s^{1/2}$<4.5 GeV; a combination of the phase space and isobar models for N+N and π+N one pion production; γA reactions extended down to the Giant Dipole Resonance energies and below; and an arbitrary light nuclear projectile (*e.g.*, d) and nuclear target (*e.g.*, d or He).

A mix-and-match is used in the transition region of $0.8 < T_1 < 4.5$ GeV to link the above explicit description and the QGSM. Here, $T_1$ is a kinetic energy of the projectile in the rest frame of the collision partner. The code considers that the nuclear reaction goes through three stages: intra-nuclear cascade, pre-equilibrium emission and evaporation/fission from the excited and thermally equilibrated residual nuclei. The nucleons with close momenta produced in the first stage can form fast light fragments (d, t, $^3$He and $^4$He) via the coalescence mechanism. Low energy neutrons, protons, nuclear fragments, de-excitation photons and fission products are generated in the last two stages. In the newest version, the extension of heavy-ion collisions to low energies (below $4V_c$, where $V_c$ is the Coulomb barrier) is done by replacing the cascade stage with formation of a compound nucleus followed by the pre-equilibrium and equilibrium evaporation/fission processes.

Results of the first comparison with the HARP data [9] of the MARS15-LAQGSM calculations of large-angle pion production on heavy nucleus are shown in Fig. 5 (left). Calculations according to the formula (2) are also shown in the Figure for this 8 GeV/c p+Pb reaction. One can see the perfect agreement with the data for negative pions with p > 200 MeV/c. At the same time, the QGSM model overestimates the data by up to 50% for lower pion momenta. Attempts to fix the problem just by increasing the pion absorption cross-section have had a limited success so far. The work is underway to improve the model in this important region.

Fig. 5 (right) shows the comparison of the MARS15-LAQGSM calculations of small- and intermediate-angle pion and proton production with the HARP data [14, 15] for the 8 GeV/c p+Cu interactions. The agreement is good for most of the momentum-angle space but in some regions the model underestimates the data by a factor of two (or even bigger for protons). It is clear that additional efforts are needed here to verify the code in complementary cases and – if needed – further improve the model in this energy domain.

Double-differential cross-sections of the 730-MeV p+Pb→$π^±$X reaction at angles from 15° to 150° are presented in Fig. 6. Keep in mind that such a low projectile energy is below the lower limit of 3 GeV for the LAQGSM use in MARS15. With this note, one can consider the agreement with data as satisfactory. The most noticeable problem in the model is still with the excessive yield of low-energy pions.

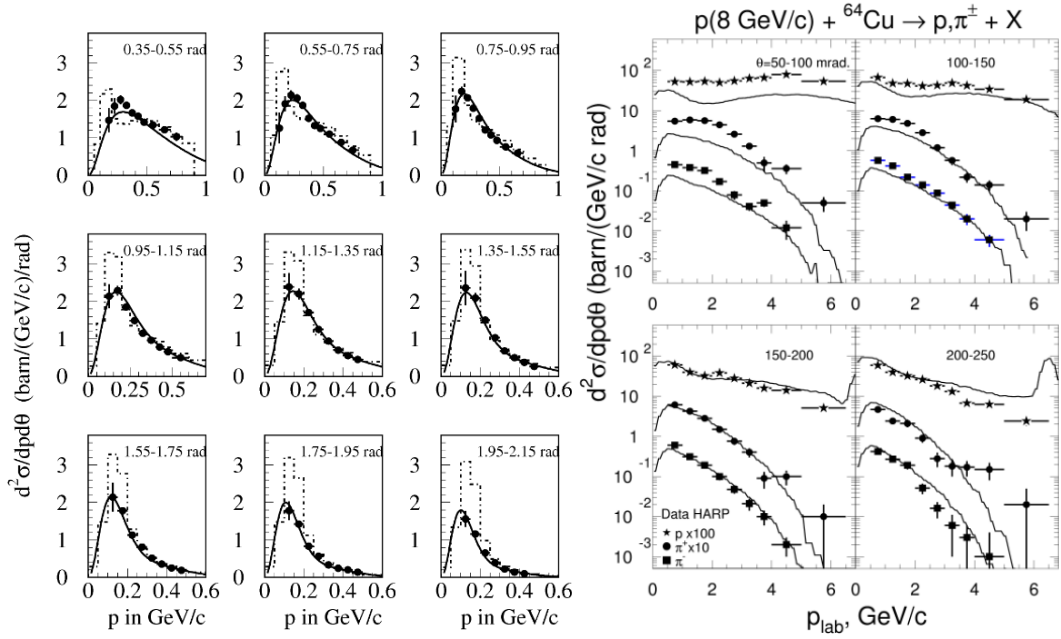

Fig. 5. Left: $\pi^-$ production in p+Pb interactions at 8 GeV/c calculated with inclusive (solid) and LAQGSM (dashed) models *vs* HARP data [9]. Right: proton and pion production in p+Cu reaction at 8 GeV/c calculated with LAQGSM model *vs* HARP data [14, 15].

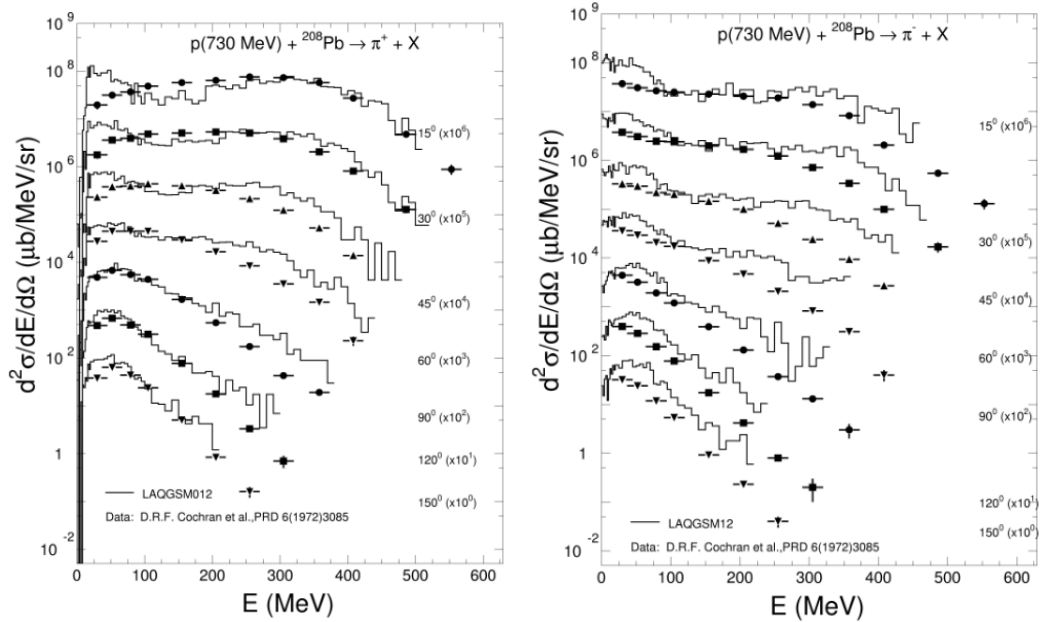

Fig. 6. Positive (left) and negative (right) pion production in p+Pb interactions at 730 MeV calculated with LAQGSM model *vs* LANL data [13].

Fig. 7 shows the double-differential cross-sections of positive and negative pions produced at four fixed angles in p+Be and p+Au interactions at 12.3 GeV/c. The MARS15-LAQGSM calculations are in a very good agreement with the BNL data [16].

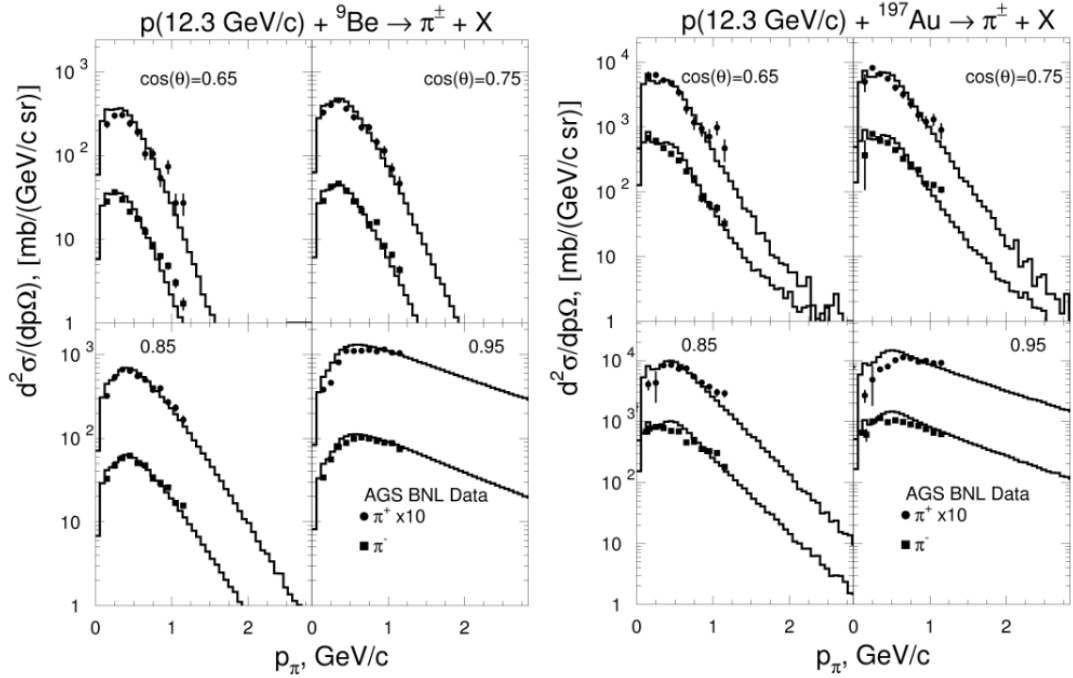

Fig.7. Pion production in p+Be (left) and p+Au (right) interactions at 12.3 GeV/c calculated with LAQGSM model *vs* data [16].

The previous comparisons in this paper were all presented for the proton-induced reactions. Double-differential cross-sections of pions and protons produced at large angles in the $\pi^-$ interactions with a lead nucleus at 3 and 5 GeV/c are shown in Fig. 8. The LAQGSM calculations are in a good agreement with the HARP-CDP data [17] for all the energies and angles.

Motivated by the plans for the next generation of rare-decay experiments in the Project X era, the MARS15-LAQGSM model for kaon production was substantially improved [18]. The proton kinetic beam energy ($T_p$) threshold for producing kaons is 1.7 GeV (on protons) and the kaon yield fraction grows with the increasing number of exclusive production channels that open and saturate around $T_p$ of 6 GeV. The efforts were specifically put on the model development in the near-threshold region. As a result, the improved model predictions were found in an excellent agreement with experimental data from the ANKE spectrometer at COSY-Julich [19] for $T_p$ near 2 GeV.

Fig. 9 (left) shows the double-differential cross-section of the 2.3-GeV p+C → $K^+$X reaction for forwardly-produced kaons. As in Ref. [18], the excellent agreement between the data and calculations done with the newest LAQGSM model in MARS15 is observed. Comparison of the new model with data [20] reveals a good agreement for $K^\pm$ large-angle production for the 3.5-GeV proton interactions with the gold nucleus.

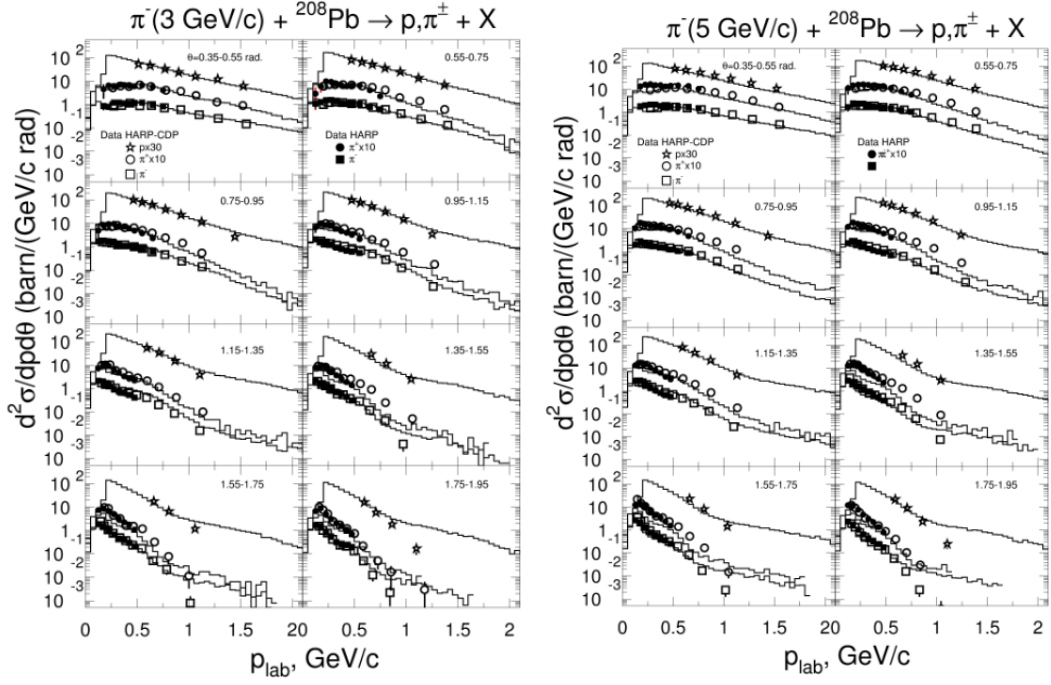

Fig.8. Proton and pion production in $\pi^-$ +Pb interactions at 3 GeV/c (left) and 5 GeV/c (right) calculated with LAQGSM model *vs* HARP-CDP data [17].

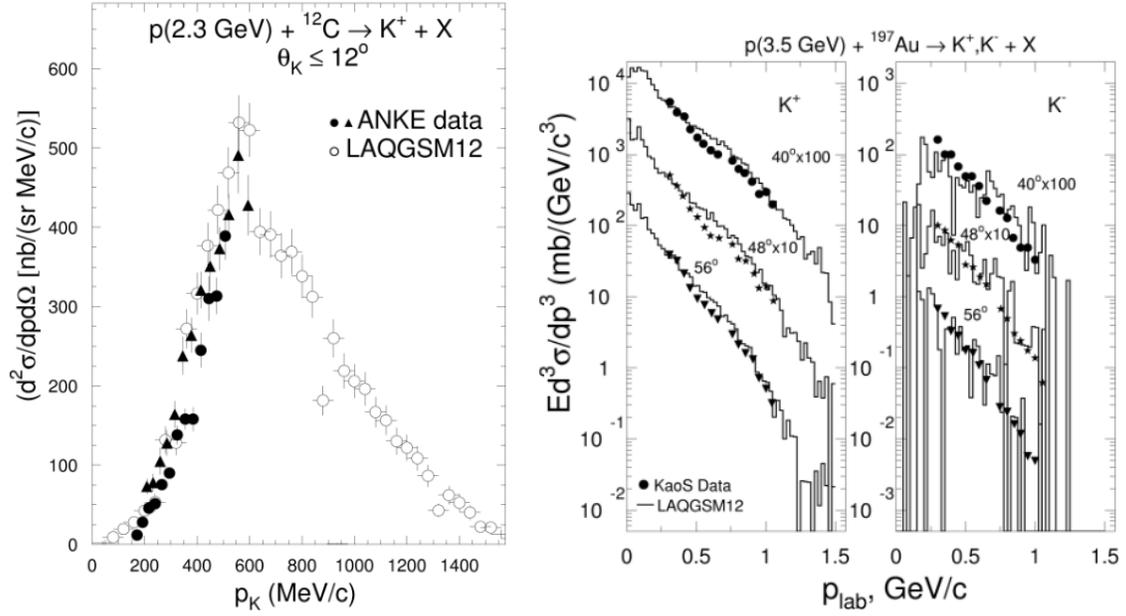

Fig.9. Kaon production in p+C interactions at 2.3 GeV (left) and p+Au interactions at 3.5 GeV (right) calculated with LAQGSM models *vs* data [19] and [20], respectively.

## Conclusions

Despite the boost the HARP experiment made to the status of data on particle yields in nuclear reactions in the 2 to 12 GeV energy range, there is still a lack of data in some phase-space regions. Moreover, in some cases there is an inconsistency between the new data and the data measured over the last decades. Also, and this is especially unfortunate, there is an inconsistency between the two HARP subgroups' data. There is certainly a noticeable progress with the theoretical models capable of an accurate prediction of particle production at intermediate energies. The phenomenological inclusive model and improved LAQGSM model described in this paper are an example. The agreement of both models with data is good for most of the momentum-angle space, but in some regions the new LAQGSM disagrees with the data by up to a factor of two. Additional efforts are still needed to further improve the model in this intermediate energy region, which is difficult from all prospects but is crucial for numerous applications.

## Acknowledgements

This work was supported by Fermi Research Alliance, LLC under contract No. DE-AC02-07CH11359 with the U.S. Department of Energy.